Comment on
"Study of visco-elastic fluid flow and heat transfer over a stretching sheet with variable viscosity", by M. Subhas Abel, Sujit Kumar Khan and K.V. Prasad [International Journal of Non-Linear Mechanics, 37, 2002, 81-88]

Asterios Pantokratoras
Associate Professor of Fluid Mechanics
School of Engineering, Democritus University of Thrace,
67100 Xanthi – Greece
e-mail:apantokr@civil.duth.gr

The above paper concerns the boundary layer flow of a visco-elastic fluid along a stretching sheet immersed in a porous medium. The plate temperature is different from that of the ambient medium and the fluid viscosity is a function of temperature while the other fluid properties are assumed to be constant. The boundary layer equations are transformed into ordinary ones and subsequently are solved using the shooting method with Runge-Kutta integration algorithm. However, there are two fundamental errors in this paper which are presented below:

1. In the transformed energy equation (2.8) the Prandtl number appears in three terms and has been assumed constant across the boundary layer. However, the Prandtl number is a function of viscosity and viscosity has been assumed a function of temperature whereas the other fluid properties are considered constant and independent of temperature. Taking into account that temperature varies across the boundary layer, the Prandtl number varies, too. The assumption of constant Prandtl number inside the boundary layer, with temperature dependent viscosity, is a wrong assumption and leads to unrealistic results (Pantokratoras, 2004, 2005, 2007). In these three paper by Pantokratoras the difference in the results between variable Prandtl number (correct assumption) and constant Prandtl number (wrong assumption) reached 435 %, 85 %.and 98 %. The problem can be treated properly either considering the Prandtl number as a variable in the transformed equations (Saikrishnan and Roy, 2003) or with the direct solution of the initial boundary layer equations and treating the viscosity as a function of temperature (Pantokratoras, 2004, 2005, 2007).
2. The viscous dissipation term in the energy equation has been modelled as

$$\Phi = \mu \left(\frac{du}{dy}\right)^2 \tag{1}$$

where μ is the fluid viscosity and u is the fluid velocity along the plate. However, in porous media the modelling of viscous dissipation is completely different from that in pure fluid flow (without porous medium). Nield (2000) proposed the following equation for modelling the viscous dissipation in a porous medium

$$\Phi = \frac{\mu u^2}{k'} - \mu u \frac{d^2 u}{dy^2} \tag{2}$$

where $k'$ is the porous medium permeability. On the other hand, Al-Hadrami et al. (2002, 2003) proposed the following equation

$$\Phi = \frac{\mu u^2}{k'} + \mu \left(\frac{du}{dy}\right)^2 \tag{3}$$

The problem is still open in the literature but one thing is sure. The viscous dissipation term in porous media is expressed by two terms and not one as happed in the above paper.

Taking into account the above arguments there are doubts for the credibility of the above work.

REFERENCES

1. Abel, M.S., Khan, S. K. and Prasad, K.V. (2002). Study of visco-elastic fluid flow and heat transfer over a stretching sheet with variable viscosity, International Journal of Non-Linear Mechanics, Vol. 37, pp. 81-88.

Comment on
Two papers published by Subhas Abel and his co-workers in International Journal of Non-Linear Mechanics and International Journal of Thermal Sciences

1. "Diffusion of chemically reactive species of non-Newtonian fluid immersed in a porous medium over a stretching sheet", by K.V. Prasad, Subhas Abel and P.S. Datti [**International Journal of Non-Linear Mechanics**, 38, 2003, pp. 651-657]

2. "Buoyancy force and thermal radiation effects in MHD boundary layer visco-elastic flow over continuously moving stretching surface", by Subhas Abel, K.V. Prasad and Ali Mahaboob [**International Journal of Thermal Sciences**, 44, 2005, pp. 465-476]

In the above papers there is a problem in some figures. It is known in boundary layer theory that velocity and temperature profiles approach the ambient fluid conditions asymptotically and do not intersect the line which represents the boundary conditions. In the following figure we show schematically one correct profile and one wrong profile.

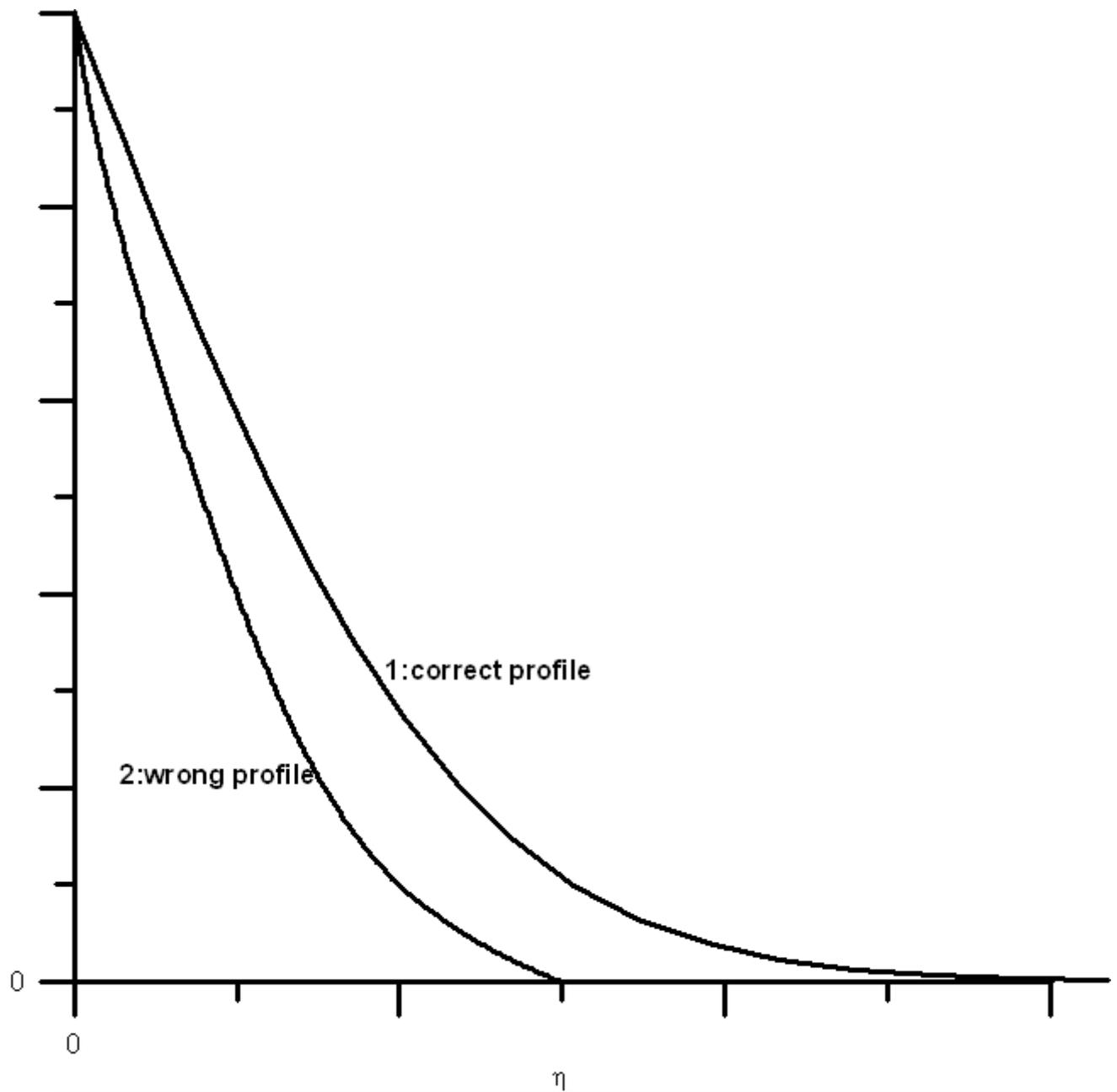

Figure1. Correct and wrong velocity or temperature profiles in boundary layer flow.

In the above papers there are many profiles which are similar to profile 2 as follows:

First paper: Two profiles in figure 1b and all profiles in figure 3. In total 9 profiles are similar to profile 2 of the above figure.

Second paper: Two profiles in figure 4a and two profiles in figure 4b. All profiles included in figure 6a. Many profiles included in figures 8a and 8b and all profiles in figure 9b. In total 28 profiles are similar to profile 2 of the above figure.

All the above profiles are probably truncated due to a small calculation domain used and are wrong.